\documentstyle[12pt,epsf,epsfig,wrapfig]{article}
\newcommand{\be}{\begin{equation}}
\newcommand{\ee}{\end{equation}}
\newcommand{\bea}{\begin{eqnarray}}
\newcommand{\eea}{\end{eqnarray}}

\newcommand{\bfk}{\mbox{\boldmath $k$}}

\newcommand{\pup}{p^\uparrow}

\newcommand{\cupar}{c^\uparrow}

\newcommand{\Aup}{A^\uparrow}

\newcommand{\la}{\lambda}
\newcommand{\hf}{\hat f}
\newcommand{\ua}{\uparrow}
\newcommand{\da}{\downarrow}
\def\lsim{\mathrel{\rlap{\lower4pt\hbox{\hskip1pt$\sim$}}\raise1pt\hbox{$<$}}}
\def\gsim{\mathrel{\rlap{\lower4pt\hbox{\hskip1pt$\sim$}}\raise1pt\hbox{$>$}}}
\def\nostrocostruttino#1\over#2{\mathrel{\mathop{\kern 0pt \rlap
{\hbox{$#1$}}} \hbox{\kern-.135em $#2$}}}

\def\lsim{\mathrel{\rlap{\lower4pt\hbox{\hskip1pt$\sim$}}\raise1pt\hbox{$<$}}}
\def\gsim{\mathrel{\rlap{\lower4pt\hbox{\hskip1pt$\sim$}}\raise1pt\hbox{$>$}}}
\def\nostrocostruttino#1\over#2{\mathrel{\mathop{\kern 0pt \rlap
{\hbox{$#1$}}} \hbox{\kern-.125em $#2$}}}

\newcommand{\ud}{\mathrm{d}}
\newcommand{\laa}{\lambda_{a}}
\newcommand{\lab}{\lambda_{b}}
\newcommand{\lac}{\lambda_{c}}
\newcommand{\lad}{\lambda_{d}}

\newcommand{\vphi}{\varphi}

\newcommand{\IoneG}{{\mathcal{T}}_{1}^{g}}
\newcommand{\ItwoG}{{\mathcal{T}}_{2}^{g}}
\newcommand{\Ione}{{\mathcal{T}}_{1}}
\newcommand{\Itwo}{{\mathcal{T}}_{2}}

\begin{document}

\begin{center}
{\bfseries GENERAL HELICITY FORMALISM FOR
SINGLE AND DOUBLE SPIN ASYMMETRIES IN $pp\to\pi+X$\footnote{Talk
  delivered by S. Melis at the ``XI Workshop on High Energy Spin Physics'', SPIN 05, September 27 - October 1, 2005, Dubna, Russia.}}

\vskip 5mm
M. Anselmino$^{1}$, M. Boglione$^{1}$, U. D'Alesio$^{2}$,
E. Leader$^{3}$, 
S. Melis$^{2}$ and
F. Murgia$^{2}$

\vskip 5mm
{\small
(1) {\it
Dipartimento di Fisica Teorica, Universit\`a di Torino and INFN, Sezione di Torino,\\
 Via P.~Giuria 1, I-10125 Torino, Italy}\\
(2) {\it
Dipartimento di Fisica, Universit\`a di Cagliari and INFN, Sezione di Cagliari,\\
C.P. 170, I-09042 Monserrato (CA), Italy}\\
(3) {\it Imperial College London, Prince Consort Road, London SW7 2BW, U.K.}
\\
}
\end{center}

\vskip 5mm
\begin{abstract}
We consider within a generalized QCD factorization approach, the high energy inclusive polarized process $pp\to\pi +X$,
including all intrinsic partonic motions. 
Several new spin and 
$\bfk_{\perp}$-dependent soft functions appear
and contribute to cross sections and
spin asymmetries. 
We present here formal expressions for transverse single spin asymmetries and double longitudinal ones. 
The transverse single spin
asymmetry, $A_N$, is considered in detail, and all contributions are
evaluated numerically. It is shown that the azimuthal phase integrations strongly suppress most contributions,
leaving at work mainly the Sivers effect.
\end{abstract}

\vskip 8mm
\section{Introduction and formalism}
In the last years we have developed an approach to study inclusive
pion production in (un)polarized hadronic collisions, at high energies
and moderately large $p_T$~\cite{fu,Anselmino:2004ky,FUMES}. 
In our approach we take into account 
the parton intrinsic motion, $\bfk_{\perp}$,
in the partonic distribution functions (PDF's), in the fragmentation sector (FF)
and in the elementary scattering processes allowing for a full non collinear kinematics.
In writing cross sections we adopt a generalization of the QCD factorization scheme.
However, at present, a factorized formula
for inclusive hadron production processes, with the inclusion of $\bfk_{\perp}$-effects,
has to be considered
as a reasonable phenomenological model,
since a corresponding factorization theorem has not been proved yet.

Let us consider the process $(A,S_A)+(B,S_B)\to C+X$, where $A$ and
$B$ are two spin one-half hadrons in a pure polarization state,
denoted respectively by $S_A$ and $S_B$. 
The cross section for this process can be given at leading order (LO) as a convolution of all possible elementary hard scattering processes, $ab\to cd$, with soft, leading twist, spin and $\bfk_{\perp}$-dependent PDF's and FF's, namely:
{\setlength\arraycolsep{2pt}\bea\label{gen1}
\ud\sigma^{(A,S_A) + (B,S_B) \to C + X}&=&\!\!\sum_{a,b,c,d, \{\la\}} \> \rho_{\la^{\,}_a,\la^{\prime}_a}^{a/A,S_A} \, \hat f_{a/A,S_A}(x_a,\bfk_{\perp a})\otimes
\rho_{\la^{\,}_b, \la^{\prime}_b}^{b/B,S_B} \,\hat f_{b/B,S_B}(x_b,\bfk_{\perp b})\nonumber\\
&&\quad\quad\quad\>\otimes\, \hat M_{\la^{\,}_c, \la^{\,}_d; \la^{\,}_a, \la^{\,}_b} \,\hat M^*_{\la^{\prime}_c, \la^{\,}_d; \la^{\prime}_a,\la^{\prime}_b}\otimes\,\hat{D}_{\la^{\,}_c,\la^{\prime}_c}^{C}(z,\bfk_{\perp C})\,.
\eea}

The notation $ \{\la\}$ implies a sum over all helicity indices,
while $a,b,c,d$ stand for a sum over all partonic contributions.
$\rho_{\la^{\,}_a,\la^{\prime}_a}^{a/A,S_A}$ and $\rho_{\la^{\,}_b, \la^{\prime}_b}^{b/B,S_B}$
are the helicity density matrices for parton $a$ and $b$ respectively,
describing the polarization state of partons inside parent hadrons.
We denote by $\hat f_{a/A,S_A}(x_a,\bfk_{\perp a})$ 
the PDF of parton $a$ inside hadron $A$, similarly for parton $b$.  
The elementary scattering process is described by means of the
helicity amplitudes $\hat M_{\lac,\lad;\laa,\lab}$ 
to be computed in the hadron c.m. frame.
Due to the non collinear partonic configuration, their expressions do not coincide anymore with 
the well known helicity amplitudes calculated in the usual canonical partonic c.m frame, denoted by $\hat M^{0}_{\lac,\lad;\laa,\lab}$.
In fact, contrary to the collinear configuration case,
the hadronic production plane does not coincide with the partonic
scattering plane. 
The amplitudes $\hat M_{\lac,\lad;\laa,\lab}$ can be computed from the $\hat M^{0}_{\lac,\lad;\laa,\lab}$ ones by means of the Lorentz transformation connecting the partonic c.m frame to the hadronic one;
this involves complicated azimuthal phases, denoted by $\vphi_i$,
which depend in a highly non trivial way on all kinematical variables
of the convolution integral in
Eq.~(\ref{gen1})~\cite{Anselmino:2004ky}. 
For massless partons there are at most three independent helicity
amplitudes for each partonic subprocess: 
\bea
&&\hat M_{++;++}  = \hat M^0_{++;++}\,e^{i\varphi_1}\equiv\hat M_1^0 \,e^{i\varphi_1} \quad\hat M_{-+;-+}  =  \hat M^0_{-+;-+}\,e^{i\varphi_2}\equiv\hat M_2^0 \,e^{i\varphi_2} \label{phases}\nonumber\\
&&\hspace{3cm}\hat M_{-+;+-}  =  \hat M^0_{-+;+-}\,e^{i\varphi_3}\equiv\hat M_3^0 \,e^{i\varphi_3}\,,
\eea
where the phases $\vphi_i$ ($i=1,2,3$) also depend on the helicity indices. 
Finally, $\hat{D}_{\la^{\,}_c,\la^{\prime}_c}^{C}(z,\bfk_{\perp C})$
is the product of the soft helicity fragmentation amplitudes for the
process $c\to C+X$. 
It can be shown that $\hat D_{C/c}(z,k_{\perp C})\equiv D_{++}^{C}(z,k_{\perp C})=D_{--}^{C}(z,k_{\perp C})$ corresponds to the usual unpolarized $k_{\perp}$-dependent FF,
while $\Delta^{N}\hat D_{C/c^{\ua}}(z,k_{\perp C})\equiv 2 \textrm{Im}
D_{+-}^C(z,k_{\perp C})$ is related to the Collins function. 
On-shell gluons cannot be transversely spin polarized;
therefore there is not a gluon Collins function.
However one can define a ``Collins-like'' function for gluons: $\Delta^{N}\hat D_{C/\IoneG}(z,k_{\perp C})\equiv 2 \textrm{Re} D_{+-}^C(z,k_{\perp C})$, related to the fragmentation of a linearly polarized gluon into an unpolarized hadron.

Eq.~(\ref{gen1}) contains all possible combinations of spin and
$\bfk_{\perp}$-dependent PDF's/FF's. 
In order to give a partonic interpretation of the spin and $\bfk_{\perp}$-dependent PDF's, we consider in detail the explicit expressions for the quark and gluon helicity density matrices.
\section{Spin and  $\bfk_{\perp}$-dependent PDF's}\label{PDF}
Let us first consider a quark parton $a$. Its helicity density matrix can be written in terms of its polarization vector components, $P_i^a$, as given in its helicity frame.
We can write:
\be\label{heldensQ}
\rho_{\la^{\,}_a, \la^{\prime}_a}^{a/A,S_A}
\>\hat f_{a/A,S_A}(x_a,\bfk_{\perp a})=\frac{1}{2}
\left(
\begin{array}{cc}
1+P^a_z & P^a_x - i P^a_y \\
 P^a_x + i P^a_y & 1-P^a_z
\end{array}
\right)\>\hat f_{a/A,S_A}(x_a,\bfk_{\perp a})\,.
\ee
For a given polarization state of hadron $A$ we can define the following functions:
\bea \label{siSZ}
P^a_i \, \hf _{a/A,S_Z}(x_a,\bfk_{\perp a}) &=& \Delta \hf_{{s_i}/S_Z}^a = 
\hf _{s_i/+}^a -  \hf _{-s_i/+}^a \equiv 
\Delta \hf_{{s_i}/+}^a(x_a, \bfk_{\perp a})\\
\label{siSY}
P^a_i \, \hf _{a/A,S_Y}(x_a,\bfk_{\perp a}) &=& \Delta \hf_{{s_i}/S_Y}^a = 
\hf _{s_i/\ua}^a -  \hf _{-s_i/\ua}^a \equiv 
\Delta \hf_{{s_i}/\ua}^a(x_a, \bfk_{\perp a})\\
\label{UnpSiv}\hf _{a/A,S_Y}(x_a, \bfk_{\perp a})& =& \hf_{a/A}(x_a, \bfk_{\perp a}) + \frac{1}{2}\,
\Delta \hf_{a/S_Y}(x_a, \bfk_{\perp a})\,,
\eea
where $S_Z$ and $S_Y$ denote respectively the direction of hadron $A$
polarization in its helicity frame.  
At LO, $\Delta \hf_{{s_i}/+}^a(x_a, \bfk_{\perp a})$ and $\Delta \hf_{{s_i}/\ua}^a(x_a, \bfk_{\perp a})$ can be easily interpreted
as the number density of partons, polarized along the $i$-direction (in the parton helicity frame), inside respectively a longitudinally or transversely polarized hadron.
$\hf _{a/A,S_Y}(x_a, \bfk_{\perp a})$ is related to unpolarized partons and can be split into two PDF's:
the usual $k_{\perp}$-dependent unpolarized PDF and
the Sivers function. 
The latter can be interpreted as the probability to find an
unpolarized parton inside a transversely polarized hadron. 
Moreover also  the function $\Delta \hf^{a}_{s_y/S_Y}(x_a, \bfk_{\perp
  a})$ can be split into two terms:
\bea\label{eq:BM}
\Delta \hf^{a}_{s_y/S_Y}(x_a, \bfk_{\perp a}) &=& \Delta \hf ^a_{s_y/A}(x_a, k_{\perp a}) +\Delta ^{-} \hf ^a_{s_y/S_Y}(x_a, \bfk_{\perp a})\\
\Delta ^- \hf ^a_{s_y/S_Y}(x_a, \bfk_{\perp a}) &\equiv &\frac 12 \, \left[
\Delta \hf ^a_{s_y/\ua}(x_a, \bfk_{\perp a}) - \Delta \hf ^a_{s_y/\da}(x_a, \bfk_{\perp a}) \right]\nonumber\,.
\eea
\noindent
$\Delta \hf ^a_{s_y/A}(x_a, k_{\perp a})$ is the
Boer-Mulders function which gives
the probability to find a transversely polarized quark inside an unpolarized hadron,
whereas $\Delta ^- \hf ^a_{s_y/S_Y}(x_a, \bfk_{\perp a})$ is related to the transversity function (see Appendix B of Ref.~\cite{FUMES})

We can also write the analogue of Eq.~(\ref{heldensQ}) for gluon partons:
\be
\rho_{\la^{\,}_a, \la^{\prime}_a}^{a/A,S_A}
\>\hat f_{a/A,S_A}(x_a,\bfk_{\perp a})=
\frac{1}{2}
\left(
\begin{array}{cc}
1+P_{z}^{g} &
\IoneG-i\ItwoG \\
\IoneG+i \ItwoG & 1-P_{z}^{g}
\end{array}
\right)\>\hat f_{a/A,S_A}(x_a,\bfk_{\perp a}) \,,
\ee
where $\IoneG$ and $\ItwoG$ are related to the Stokes parameters for
linearly polarized gluons. 
Similarly to the quark case we can define the functions:
\bea
\label{I12SY}
\IoneG \, \hf _{g/A,S_Y}(x_a,\bfk_{\perp a})\equiv
\Delta \hf_{{\Ione}/\ua}^g(x_a,\bfk_{\perp a}) &&
\ItwoG \, \hf _{g/A,S_Y}(x_a,\bfk_{\perp a})\equiv
\Delta \hf_{{\Itwo}/\ua}^g(x_a,\bfk_{\perp a})\\
\label{I12SZ}
\IoneG \, \hf _{g/A,S_Z}(x_a,\bfk_{\perp a})\equiv
\Delta \hf_{{\Ione}/+}^g(x_a,\bfk_{\perp a})&&
\ItwoG  \, \hf _{g/A,S_Z}(x_a,\bfk_{\perp a})\equiv
\Delta \hf_{{\Itwo}/+}^g(x_a,\bfk_{\perp a})\\
\label{PZGYZ}
P^g_z \, \hf _{g/A,S_Y}(x_a,\bfk_{\perp a})\equiv
\Delta \hf_{s_z/\ua}^g(x_a,\bfk_{\perp a})&&
P^g_z \, \hf _{g/A,S_Z}(x_a,\bfk_{\perp a})\equiv
\Delta \hf_{s_z/+}^g(x_a,\bfk_{\perp a})\,.
\eea
Eqs.~(\ref{I12SY}) and (\ref{I12SZ}) are related respectively to the probability to find linearly polarized gluons inside a transversely or longitudinally polarized hadron $A$.
Eq.~(\ref{UnpSiv}) holds also for unpolarized gluons.
Furthermore, in analogy to Eq.~(\ref{eq:BM}) we can write:
\bea
\Delta \hat f^g_{\Ione/\uparrow}(x_a,\bfk_{\perp a})&=&
\Delta \hat f^g_{\Ione/A}(x_a,k_{\perp a})+\Delta^-\hat f^g_{\Ione/\uparrow}(x_a,\bfk_{\perp a})\\
\Delta^-\hat f^g_{\Ione/\uparrow}(x_a,\bfk_{\perp a}) &=& \frac 12 \, \left[
\Delta \hat f^g_{\Ione/\uparrow}(x_a,\bfk_{\perp a}) -
\Delta \hat f^g_{\Ione/\downarrow}(x_a,\bfk_{\perp a}) \right]\,,\nonumber 
\eea
where $\Delta \hat f^g_{\Ione/A}(x_a,k_{\perp a})$ is the analogue of
the Boer-Mulders function and is related to 
linearly polarized gluons inside an unpolarized hadron. $\Delta^-\hat f^g_{\Ione/\uparrow}$ is the analogue of the transversity and is related to linearly polarized gluons inside a transversely polarized hadron. 
\section{Spin Asymmetries}
From Eq.~(\ref{gen1}) and using the PDF's defined in Sec.~\ref{PDF}, we can calculate unpolarized cross sections, single (SSA) and double (DSA) spin asymmetries.
As an application, we consider the numerator of the transverse SSA for
the process $A^{\uparrow}B\to C+X$, limiting ourselves to show the
results only for two partonic subprocesses: $q_a q_b \to q_c q_d$ and
$g_a g_b \to g_c g_d$. For the  $q_a q_b \to q_c q_d$ contribution we
have:  
\bea
&&\hspace*{-1.3cm}[d\sigma(A^{\uparrow}B\to C+X)-d\sigma(A^{\downarrow}B\to C+X)]^{q_a q_b \to q_c q_d}\propto
\nonumber \\
&\,& \frac{1}{2} \, \Delta \hf_{a/\Aup} (x_a,\bfk_{\perp a}) \,
\hf_{b/B}(x_b, k_{\perp b}) \,
\left[\,|{\hat M}_1^0|^2 + |{\hat M}_2^0|^2 + |{\hat M}_3^0|^2 \right] \,
\hat D _{C/c} (z, k_{\perp C}) \nonumber \\
&+&  2\,\left[ \Delta^- \hf^a_{s_y/\ua} (x_a, \bfk_{\perp a}) \, 
\cos(\varphi_3 -\varphi_2) -\Delta \hf^a_{s_x/\ua} (x_a,\bfk_{\perp a}) \, 
\sin(\varphi_3 -\varphi_2) \right] \,
\nonumber \\
&& \times \, \Delta \hf^b_{s_y/B}(x_b,\bfk_{\perp b})\,
{\hat M}_2^0 \, {\hat M}_3^0 \,\hat D _{C/c} (z, k_{\perp C})\nonumber \\
&+& \left[ \Delta^- \hf^a_{s_y/\ua} (x_a, \bfk_{\perp a})\,
\cos(\varphi_1 -\varphi_2 + \phi_C^H)- \Delta \hf^a_{s_x/\ua}(x_a,\bfk_{\perp a})\,
\sin(\varphi_1 -\varphi_2 + \phi_C^H) \right] \,\nonumber \\
&&\times \, \hf_{b/B}(x_b, k_{\perp b})\,
{\hat M}_1^0 \, {\hat M}_2^0 \, \Delta^N {\hat D}_{C/\cupar} (z, k_{\perp C})\,\label{num-asym-qq}
 \\
&+&  \frac{1}{2} \, \Delta \hf_{a/\Aup} (x_a, \bfk_{\perp a})\,
\Delta \hf^b_{s_y/B} (x_b, \bfk_{\perp b}) \, 
{\hat M}_1^0 \, {\hat M}_3^0 \,
\Delta ^N {\hat D} _{C/\cupar} (z, k_{\perp C})\cos(\varphi_1 -\varphi_3 + \phi_C^H)\nonumber\,, 
\eea
whereas for the $g_a g_b \to g_c g_d$ contribution we have:
\bea
&&\hspace*{-1.3cm}[d\sigma(A^{\uparrow}B\to C+X)-d\sigma(A^{\downarrow}B\to C+X)]^{g_a g_b\to g_c g_d}\propto
\nonumber \\
&\,& \frac{1}{2} \, \Delta \hf_{g/\Aup} (x_{a}, \bfk_{\perp a}) \,
\hf_{g/B}(x_{b}, k_{\perp b}) \,
\left[ \, |{\hat M}_1^0|^2 + |{\hat M}_2^0|^2 + |{\hat M}_3^0|^2 \right] \,
\hat D _{C/g} (z, k_{\perp C}) \nonumber \\
&+&  2 \, \left[ \Delta^- \hf^g_{\Ione/\ua} (x_{a}, \bfk_{\perp a}) \,
\cos(\varphi_3 -\varphi_2) +\Delta \hf^g_{\Itwo/\ua} (x_{a}, \bfk_{\perp a}) 
\, \sin(\varphi_3 -\varphi_2) \right] \, \nonumber \\
&& \times \, \Delta \hf^g_{\Ione/B}(x_b,\bfk_{\perp b})\,
{\hat M}_2^0 \, {\hat M}_3^0 \,\hat D _{C/g} (z, k_{\perp C}) \nonumber
 \\
&+& \> \left[ \Delta^- \hf^g_{\Ione/\ua} (x_{a}, \bfk_{\perp a})\,
\cos(\varphi_1 -\varphi_2 + 2 \phi_C^H)+ \Delta \hf^g_{\Itwo/\ua} (x_{a}, \bfk_{\perp a})\,
\sin(\varphi_1 -\varphi_2 + 2\phi_C^H) \right] \nonumber \\
&& \times \, \hf_{g/B}(x_{b}, k_{\perp b})\,
{\hat M}_1^0 \, {\hat M}_2^0 \, \Delta^N \hat{D}_{C/\IoneG}(z, k_{\perp C})\label{num-asym-gg}
 \\
&+& \frac 12 \, \Delta \hf_{g/\Aup} (x_{a},\bfk_{\perp a})\,
\Delta \hf^g_{\Ione/B}(x_{b}, \bfk_{\perp b})\,
{\hat M}_1^0 \, {\hat M}_3^0 \,
\Delta^N \hat{D}_{C/\IoneG}(z, k_{\perp C})\cos(\varphi_1 -\varphi_3 + 2\phi_C^H) \,.\nonumber
\eea

\begin{wrapfigure}[30]{R}{6.5cm}
\begin{center}
\mbox{\epsfig{figure=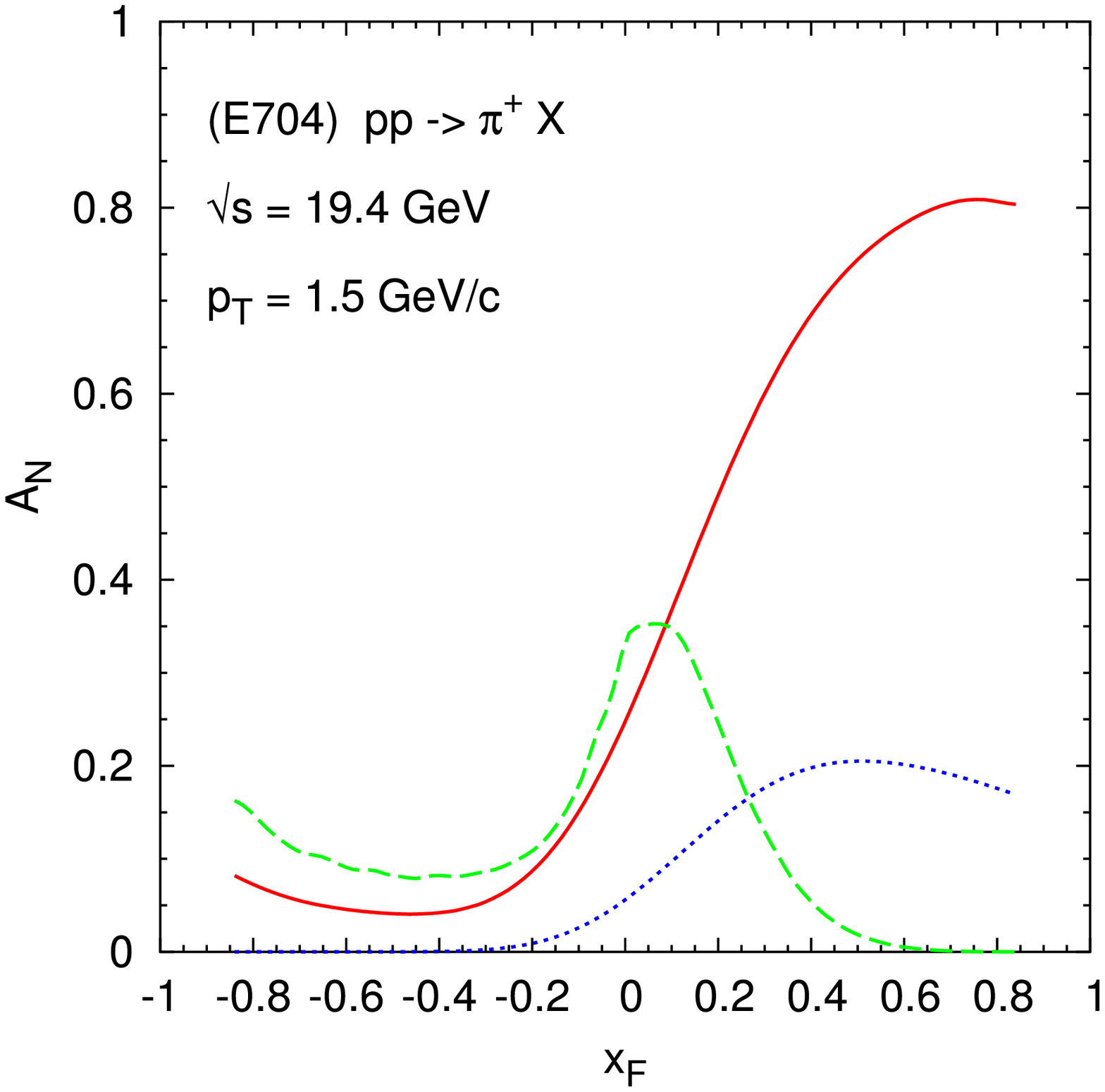,angle =-90,width=5.8cm}}
\end{center}\label{ssaE704_fig1}
{\small{\bf Figure 1.}
Different maximized contributions to $A_N$, for the process 
$\pup p \to \pi^+ \, X$ and E704 kinematics,
plotted as a 
function of $x_F$.
The different curves correspond to: {\it solid line} = quark Sivers 
mechanism; {\it dashed line} = gluon Sivers mechanism; 
{\it dotted line} = transversity $\otimes$ Collins. All other contributions
are much smaller. }
\end{wrapfigure}

Let us briefly comment on Eqs.~(\ref{num-asym-qq}) and (\ref{num-asym-gg}).
There are four terms contributing to the numerator of the transverse SSA:
the Sivers contribution (1st term); the transversity (``transversity
like''in Eq.~(\ref{num-asym-gg})) $\otimes$ Boer-Mulders (``like'')
contribution (2nd term); the transversity (``like'') $\otimes$ Collins
(``like'') contribution (3rd term); the Sivers $\otimes$ Boer-Mulders
(``like'') $\otimes$ Collins (``like'') contribution (4th term). 
The term ``like'' is referred to those functions in Eq.~(\ref{num-asym-gg}),
presented in Sec.~\ref{PDF}, which are related to linearly polarized gluons and are the analogues of those for transversely polarized quarks in Eq.~(\ref{num-asym-qq}).

We have performed a numerical evaluation of the maximal contribution of each (unknown) term appearing in Eqs.~(\ref{num-asym-qq}) and (\ref{num-asym-gg}) for the transverse SSA in $p^{\ua}p\to \pi+X$.
To this purpose, we saturate the positivity bounds for the Sivers and Collins functions and replace all other spin and $\bfk_{\perp}$-dependent PDF's with the corresponding unpolarized ones.
We sum all possible contributions with the same sign.
We assume a Gaussian $\bfk_{\perp}$ dependence for PDF's with $\langle
k_{\perp}^{2}\rangle^{1/2}=0.8$ GeV/$c$ while for FF's $\langle
k_{\perp C}^{2}(z)\rangle^{1/2}$ is taken as in Ref.~\cite{fu}. For the unpolarized PDF's and FF we adopt the MRST01~\cite{MRST01} and KKP~\cite{KKP} sets.
For more detail see Refs.~\cite{fu,FUMES}.
As an example, we consider here the evaluation of several contributions for the process $pp\to\pi^{+}+X$ in the E704 kinematical regime, $\sqrt{s}=19.4$ GeV and $p_T=1.5$ GeV/$c$, Fig.~1. We can see that the Sivers effect is the dominant contribution. The second most relevant contribution is the transversity $\otimes$ Collins contribution. All other contributions give asymmetries compatible with zero.
Such a result can be understood by looking for instance at  Eqs.~(\ref{num-asym-qq}) and (\ref{num-asym-gg}). All contributions are coupled with complicate non trivial azimuthal phases.
Under phase space integration these azimuthal phases kinematically suppress some contributions.
Notice that from our calculations the Sivers effect alone can explain, in principle, the observed SSA while the Collins effect cannot.
Our approach applies also to DSA. Let us consider as an example the $q_aq_b\to q_cq_d$ contribution to the longitudinal  DSA, $A_{LL}$:
\bea
&&\hspace*{-1.cm}[d\sigma(A^{+}B^{+}\to C+X)-d\sigma(A^{+}B^{-}\to C+X)]^{q_aq_b\to q_cq_d}\propto
\nonumber \\
&&\Delta \hat{f}^{a}_{s_z/+}(x_a, k_{\perp a})\,\Delta\hat{f}^{b}_{s_z/+}(x_b,k_{\perp b})\,\left[ \, |{\hat M}_1^0|^2 - |{\hat M}_2^0|^2 - |{\hat M}_3^0|^2 \right]\,\hat{D}_{C/c}(z,k_{\perp C})\nonumber\\
&+&2 \,\Delta \hat{f}^{b}_{s_x/+}(x_b, k_{\perp b}){\hat M}_2^0 \, {\hat M}_3^0\,\hat{D}_{C/c}(z,k_{\perp C})\times\nonumber\\
&&\quad\left[\,\Delta \hat{f}^{a}_{s_x/+}(x_a, k_{\perp
    a})\,\cos(\vphi_3-\vphi_2) 
+\Delta \hat{f}^{a}_{s_y/A}(x_a, k_{\perp a})\,\sin(\vphi_3-\vphi_2)\right]\label{all-qqqq}\\
&-&\hf_{a/A}(x_a, k_{\perp a})\,\Delta\hat{f}^{b}_{s_x/+}(x_b,k_{\perp b})\,{\hat M}_1^0 \, {\hat M}_3^0\,\sin(\varphi_1 -\varphi_3 +\phi_C^H)\,\Delta ^N {\hat D} _{C/\cupar} (z, k_{\perp C})\,.\nonumber
\eea

The 2nd line of Eq.~(\ref{all-qqqq}) corresponds to the usual contribution coming from the helicity PDF's. The other terms come from contributions of combinations of transversely polarized quarks both in the PDF/FF sectors.
Similarly
for gluon channels with contributions of linearly polarized gluons.

Our formalism can also be applied to the study of inclusive $\Lambda$ production in hadronic collisions, Drell-Yan and SIDIS processes~\cite{Anselmino:2005ea}. As a further application, a numerical evaluation of maximal contribution to $A_{LL}$ is in progress.


\begin{thebibliography}{99}
\bibitem{fu}
U.~D'Alesio and F.~Murgia, Phys. Rev. D{\bfseries 70}, 074009 (2004)
\bibitem{Anselmino:2004ky}
M.~Anselmino, M.~Boglione, E.~Leader, U.~D'Alesio and F.~Murgia, 
Phys. Rev. D{\bfseries 71}, 014002 (2005)
\bibitem{FUMES}
M.~Anselmino, M.~Boglione, E.~Leader, U.~D'Alesio, S.~Melis and
F.~Murgia, 
Phys. Rev. D{\bfseries 73}, 014020 (2006)
\bibitem{Anselmino:2005ea}
M.~Anselmino, M.~Boglione, U.~D'Alesio A.~Kotzinian, F.~Murgia and A.~Prokudin, Phys. Rev. D{\bfseries 72},094007 (2005)
\bibitem{MRST01}
A.D.~Martin, R.G.~Roberts, W.J.~Stirling and R.S.~Thorne, Phys Lett. B{\bfseries 531}, 216 (2002)
\bibitem{KKP}
B.A.~Kniehl, G.~Kramer, B.~Potter, Nucl. Phys. B{\bfseries 582}, 514 (2000)
\end{thebibliography}
\end{document}